\documentclass[twocolumn]{aastex62}
\usepackage{bm}
\usepackage{fourier}
\usepackage{wasysym}
\usepackage{multirow}

\usepackage[caption=false]{subfig}

\usepackage{hyperref}
\usepackage{natbib}
\newcommand{\Rsub}{R_{\rm sub}}

\shortauthors{Kishimoto, Anderson, ten Brummelaar et al.}
\shorttitle{Dust sublimation region of NGC4151 as resolved by the CHARA Array}

\begin{document}
\title{
The dust sublimation region of the Type 1 AGN NGC4151 at a hundred micro-arcsecond scale \\
as resolved by the CHARA Array interferometer
}

\correspondingauthor{Makoto Kishimoto}
\email{mak@cc.kyoto-su.ac.jp}

\author{Makoto Kishimoto}
\affiliation{Department of Astrophysics \& Atmospheric Sciences, Kyoto Sangyo University, Kamigamo-motoyama, Kita-ku, Kyoto 603-8555, Japan}

\author{Matt Anderson}
\affiliation{Georgia State University Center for High Angular Resolution Astronomy Array, Mount Wilson Observatory, USA}

\author{Theo ten Brummelaar}
\affiliation{Georgia State University Center for High Angular Resolution Astronomy Array, Mount Wilson Observatory, USA}

\author{Christopher Farrington}
\affiliation{Georgia State University Center for High Angular Resolution Astronomy Array, Mount Wilson Observatory, USA}

\author{Robert Antonucci}
\affiliation{Department of Physics, University of California, Santa Barbara, USA}

\author{Sebastian Hoenig}
\affiliation{Department of Physics \& Astronomy, University of Southampton, UK}

\author{Florentin Millour}
\affiliation{Laboratoire Lagrange, Universit\'e C\^ote d'Azur, Observatoire de la C\^ote d'Azur, CNRS, Nice, France}

\author{Konrad Tristram}
\affiliation{European Southern Observatory, Santiago, Chile}

\author{Gerd Weigelt}
\affiliation{Max Planck Institute for Radio Astronomy, Bonn, Germany}

\author{Laszlo Sturmann}
\affiliation{Georgia State University Center for High Angular Resolution Astronomy Array, Mount Wilson Observatory, USA}

\author{Judit Sturmann}
\affiliation{Georgia State University Center for High Angular Resolution Astronomy Array, Mount Wilson Observatory, USA}

\author{Gail Schaefer}
\affiliation{Georgia State University Center for High Angular Resolution Astronomy Array, Mount Wilson Observatory, USA}

\author{Nic Scott}
\affiliation{Georgia State University Center for High Angular Resolution Astronomy Array, Mount Wilson Observatory, USA}


\begin{abstract}

The nuclear region of Type 1 AGNs has only been partially resolved so far in the near-infrared (IR) where we expect to see the dust sublimation region and the nucleus directly without obscuration. Here we present the near-IR interferometric observation of the brightest Type 1 AGN NGC4151 at long baselines of $\sim$250 m using the CHARA Array, reaching structures at hundred micro-arcsecond scales. The squared visibilities decrease down to as low as $\sim$0.25, definitely showing that the structure is resolved. Furthermore, combining with the previous visibility measurements at shorter baselines but at different position angles, we show that the structure is elongated {\it perpendicular} to the polar axis of the nucleus, as defined by optical polarization and a linear radio jet. A thin-ring fit gives a minor/major axis ratio of $\sim$0.7 at a radius $\sim$0.5 mas ($\sim$0.03 pc). This is consistent with the case where the sublimating dust grains are distributed preferentially in an equatorial plane in a ring-like geometry, viewed at an inclination angle of $\sim$40\degr. Recent mid-IR interferometric finding of polar-elongated geometry at a pc scale, together with a larger-scale polar outflow as spectrally resolved by the HST, would generally suggest a dusty, conical and hollow outflow being launched presumably in the dust sublimation region. This might potentially lead to a polar-elongated morphology in the near-IR, as opposed to the results here. We discuss a possible scenario where an episodic, one-off anisotropic acceleration formed a polar-fast and equatorially-slow velocity distribution, having lead to an effectively flaring geometry as we observe.

\end{abstract}

\keywords{Active galactic nuclei --- Interferometry}

\section{Introduction}\label{sec_intro}

With long-baseline infrared interferometry, it is possible to spatially resolve the thermal emission from nuclear structures on sub-parsec scales in the nuclei of active galaxies (Active Galactic Nuclei, or AGN). Such AGN are thought to comprise a supermassive black hole surrounded by optically thick material that obscures sight of the central putative accretion disk and ionized gas if viewed from equatorial direction, while this nucleus is directly seen along polar direction \citep{Antonucci85}. 
The latter polar-viewed objects are believed to be Type 1s, while the former equatorially-viewed ones are thought to be Type 2s. The system of this optically-thick, dusty material has been called a ’torus’. 

Mid-infrared (mid-IR) interferometry explores the relatively outer region of the torus where the temperature of dust grains heated by the nucleus is $\sim$300 K, while the interferometry in the near-IR can probe the innermost dusty structure where dust grains are almost at the sublimation temperature of $\sim$1500 K. Here in this paper, we focus on this innermost region. For its exploration, the most suitable objects would be Type 1 AGNs, where the innermost region is directly seen without significant obscuration and without much complication and confusion from dust extinction.

However, the overall angular size of the Type 1 nuclei turned out to be quite small. The innermost dusty region, or the dust sublimation region, is expected to be larger for brighter objects (the angular size would roughly be proportional to the square-root of apparent flux; e.g. Figure~5 of \citealt{Kishimoto07}), but even in the brightest Type 1 source on the sky, NGC4151, the region is only partially resolved with the Keck interferometer’s (hereafter KI) 85 m baseline (i.e. shows only a small decrease of visibility; \citealt{Swain03,Kishimoto09KI,Pott10}). Other Type 1 AGNs are also only partially resolved with Very Large Telescope Interferometer (VLTI) baselines up to 130~m (\citealt{Weigelt12,Dexter19,Leftley21}). 

Another way to measure the radius of the dust sublimation region is to detect the time lag of the near-IR emission variation with respect to that of the nuclear optical emission (\citealt{Koshida09} and references therein), which has been very successful. The interferometric radii inferred from the small visibility decrease described above are in fact consistent with the time-lag radius where the interferometric radii, which are brightness-weighted radii giving the overall size of the dust sublimation region, are slightly larger than the near-IR reverberation radii, which are response-weighted providing the inner boundary radius (e.g. \citealt{Kishimoto09KI,Kishimoto11}; for NGC4151, the interferometric radius is $\sim$0.5 mas, while the time-lag radius corresponds to $\sim$0.3 mas; see Figures~3 and 4 in \citealt{Kishimoto13KI}).
If this interferometric size estimate is really correct, the observations with much longer baselines would reveal a large decrease of visibility, showing a definitely resolved structure.
Here we present observations of NGC4151 with the CHARA Array at $\sim$250 m baselines, a factor of $\sim$2 longer than previously achieved for AGNs, or any other extragalactic objects in fact. We show that the near-IR visibility goes down significantly at this long baseline and argue that we started to see morphological information on this target.

\section{Observations}

We have observed NGC4151 with the CHARA Array \citep{tenBrummelaar05} over several different nights as summarized in Table~\ref{tab_data}. We used two baselines -- a short, $\sim$34 m baseline using the telescopes S1 and S2 in 2020, and a long, $\sim$250 m baseline with telescopes S2 and W1 in 2021. In all occasions, the telescopes were equipped with an adaptive optics (AO) system, which has been commissioned recently. 
The details and operations of the CHARA Array AO system are described by 
\cite{tenBrummelaar18} and \cite{Anugu20}, while the early design of the AO system is described by \cite{Che13}.
We used the instrument CLASSIC, a two-beam combiner which is the most sensitive to faint targets among the various instruments at the CHARA Array. We used three different calibrators listed in Table~\ref{tab_calibrator}, which are very close in projected distance to NGC4151. They vary in optical and infrared magnitudes, with TIC9453648 closest in optical magnitude to NGC4151. The science target was bracketed by calibrator observations close in time (except for the observation on 2020-02-15 when the time interval for one of the calibrator observations was long; see Appendix). We observed them all in K’-band ($\lambda_{\rm obs}$ = 2.13 $\mu$m), not spectrally dispersed. The details of the sequences and raw visibilities observed are shown in Appendix.

\begin{table*}[ht!]
\centering  
 \caption{Observed visibilities for NGC4151 at CHARA/CLASSIC in K’ band}
 \begin{tabular}{l c c c c c c c c}
  \hline \hline
  Observing & Time & telescopes & projected    & baseline   & $V^2$ & $V^2$ corr. for \\
  Date (UT) & (UT) &            & baseline (m) & PA (\degr) &       & $f_{\rm AD}$=0.16$\pm$0.05\\
  \hline
  2020-02-15 & 09h21m & S1-S2 &  33.9 & 179.0 & 0.87 $\pm$ 0.15 & 0.85 $\pm$ 0.17\\
  2020-02-16 & 09h21m & S1-S2 &  33.9 &   7.3 & 1.00 $\pm$ 0.27 & 1.00 $\pm$ 0.33\\
  2021-03-19 & 07h44m & S2-W1 & 249.1 & 142.2 & 0.25 $\pm$ 0.08 & 0.17 $\pm$ 0.08\\
  2021-04-30 & 05h18m & S2-W1 & 249.4 & 139.1 & 0.36 $\pm$ 0.09 & 0.28 $\pm$ 0.10\\
  2021-04-30 & 08h30m & S2-W1 & 216.6 & 113.3 & 0.63 $\pm$ 0.07 & 0.57 $\pm$ 0.08\\
  2021-04-30 & 09h23m & S2-W1 & 193.2 & 107.4 & 0.57 $\pm$ 0.15 & 0.50 $\pm$ 0.16\\
  \hline
 \end{tabular}
 \label{tab_data}
\end{table*}

\begin{table}[ht!]
 \caption{Calibrators and projected distance $r_{\rm proj}$ from NGC4151.}
 \begin{tabular}{l c c c c}
  \hline \hline
  name & V (mag) & K (mag) & $r_{\rm proj}$ (\degr) & diameter (mas)$^a$\\
  \hline
  HD105881   &  8.0 & 7.0 & 0.3  & 0.161 $\pm$ 0.004 \\
  SAO62878   &  9.8 & 8.3 & 0.09 & 0.092 $\pm$ 0.002 \\
  TIC9453648 & 11.6 & 8.7 & 0.2  & 0.090 $\pm$ 0.002 \\
  \hline
 \end{tabular}
 \label{tab_calibrator}
 $^a$ Diameters from SearchCal by JMMC.
\end{table}

\section{Data processing}

The data were processed following the steps described by \cite{tenBrummelaar13}. The instrument CLASSIC scans each fringe with a dithering mirror. For NGC4151 and all calibrators, we needed to scan relatively slowly, taking about 1.5 seconds to scan over $\sim$70 $\lambda_{\rm obs}$ (scanning frequency of $\sim$250 Hz, 5 samples per fringe, scanning over 150 $\mu$m). After subtracting an averaged sky scan (scan of nearby sky region which is regarded as a ‘dark’ frame) from each scan, we first normalize each target scan by its low-pass-filtered version. Then we Fourier-transform the normalized scan to measure its fringe power by computing the power spectrum. The noise bias has been subtracted using the average power spectrum of the off-fringe scans. 

We have constructed our own IDL scripts to implement these steps, and calculated 10-scan, or 15-second, averages of squared visibilities $V^2$ over each fringe track which lasted about 10 minutes (5 minutes on fringe + 5 minutes off fringe). This confirmed overall stability of the fringes, but also clearly identified certain tracks where fringes were within the scan range only in some portion of the track. In these cases, we only used one continuous range per track where fringes are visible in the $V^2$ track. Then we integrated all the power spectra from each track (or the designated range of the track) to calculate the final raw $V^2$ and also its uncertainty from the fluctuation of the raw $V^2$ over the track. The fluctuation of the off-fringe scans is also calculated and added in quadrature to the raw $V^2$ uncertainty. The system visibility for each observation of NGC4151 was determined from the adjacent measurements of calibrators. Finally we obtained the final calibrated visibility by dividing the raw $V^2$ with the estimated system visibility. The details of these processing and raw $V^2$ tracks are shown in Appendix.

\begin{figure}[t!]
\includegraphics[width=\columnwidth]{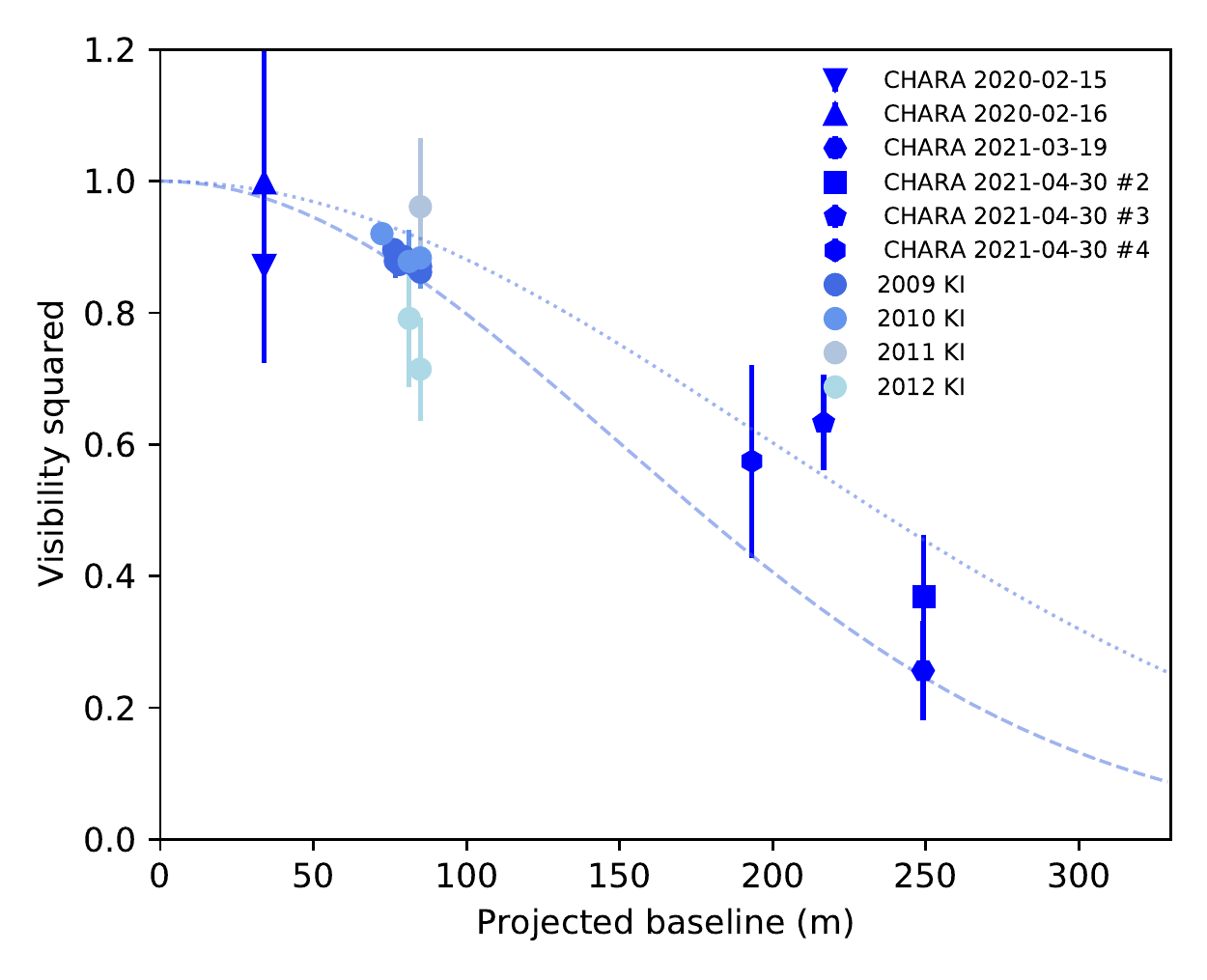}
\caption{Observed visibility squared $V^2$ as a function of projected baselines. In addition to our CHARA Array data at $\sim$30 m and $\sim$200$-$250 m baselines, we also plotted visibilities observed at Keck interferometer (\citealt{Kishimoto13KI} and references therein). $V^2$ for gaussian with HWHM of 0.3 and 0.4 mas are also plotted for reference in dotted and dashed curves, respectively.}
\label{fig_vis1d}
\end{figure}

\begin{figure}[t!]
\includegraphics[width=\columnwidth]{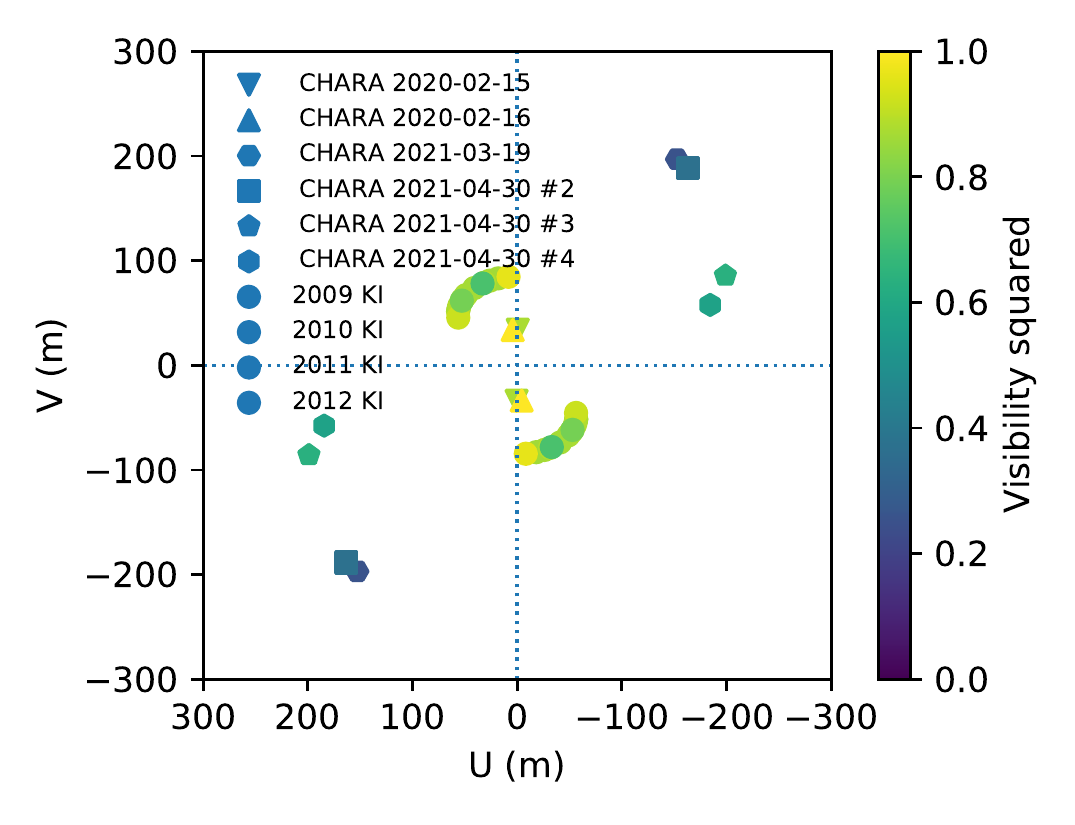}
\caption{Sampled uv points, color-coded with observed visibility squared.}
\label{fig_vis2d}
\end{figure}

\section{Results}\label{sec_res}

\subsection{Visibility vs baseline}

Figure~\ref{fig_vis1d} shows all the calibrated squared visibilities $V^2$ measured at short ($\sim$34 m) and long projected baselines ($\sim$190$-$250 m) with the CHARA Array, together with the previous measurements using the Keck interferometer (KI; \citealt{Swain03,Kishimoto09KI,Pott10,Kishimoto11,Kishimoto13KI}).
\footnote{We have VLTI/AMBER data for NGC4151 taken in 2012 as shown in \cite{Kishimoto13KI}, but we do not include them in our analysis here, since the VLTI/AMBER data might be affected by the coherence loss due to AO performance which we now fear has not been well quantified. However, the radius measurement for 2012 is represented by our KI measurements in 2012, which we include here. The radius implied by the combination of KI and VLTI data is consistent with the one by KI, as quantified in Table~1 of \cite{Kishimoto13KI}.}
The KI measurements only showed a hint of $V^2$ decrease as a function of baselines with only a small range of projected baseline lengths. Thus the data did not rule out the case where there is an unresolved source of a high flux fraction ($\sim$90\%) with the rest of the flux coming from a large scale structure. On the other hand, for different Type 1 AGNs, VLTI instruments AMBER and GRAVITY have shown that the visibilities decrease slightly over a wider range of baselines from $\sim$40~m to 130~m (\citealt{Weigelt12,Dexter19}). Here we show that the visibility does actually decrease significantly at long baselines in NGC4151, establishing that we are definitely resolving the structure spatially.

\subsection{Position angle dependence}

The CHARA Array measurements reported here and the existing KI measurements for NGC4151 are in fact complementary in the coverage of the position angle (PA) of the projected baselines. Figure~\ref{fig_vis2d} shows the uv coverage of all the observations, color-coded with the measured $V^2$. 

\begin{figure}[t]
\vspace{-1mm}%
\includegraphics[width=\columnwidth]{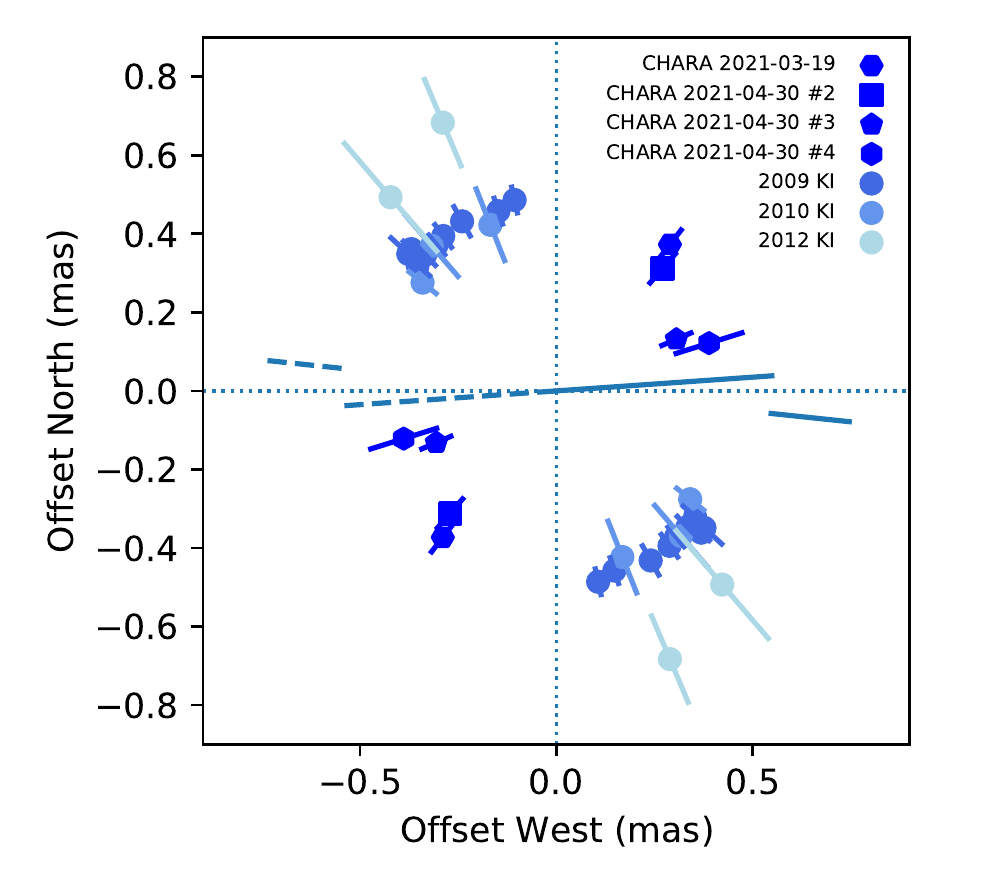}
\caption{Thin-ring radius implied by each visibility observed (after the small correction for the putative AD component), plotted along the PA of the corresponding baseline. The PA of the optical polarization (94\degr), indicative of the system polar axis, is shown as the inner dashed-solid line, while the PA of the linear radio jet structure (84\degr; components D1/D2/D3 over 50 mas scale; \citealt{Mundell03}) is indicated with the outer dashed-solid line segments. }
\label{fig_radius}
\end{figure}

Firstly, the four data points at 190$-$250~m baselines obtained with the CHARA Array actually suggest a dependency of $V^2$ on the PA, in the sense that the structure seems to be a bit more compact at PA$\sim$110\degr\ than at $\sim$140\degr. To further investigate this, we show in Figure~\ref{fig_radius} the implied thin-ring radius for each $V^2$ measurement along the PA of the baselines for the CHARA Array measurements, as well as the KI measurements (see more below for the corrections implemented). The three $V^2$ data points which are consistent with unity within uncertainties are not shown in this plot, since they do not give meaningful radius constraints alone. As a whole, we seem to see a moderate elongation along roughly a north-south direction. 


Here we have chosen to use a thin-ring model rather than a Gaussian geometry, since the central part is considered devoid of dust emission due to dust sublimation (see more discussions in section~\ref{sec_whyeq}). We are still unable to differentiate these geometries from the data themselves which are within the first lobe of the visibility curve. The ring is adopted to be infinitely thin to minimize the number of parameters involved.

Furthermore, we do expect that the central part, even if devoid of dust emission, has another thermal emission component from the putative accretion disk (AD). Based on the optical to near-IR spectral energy distribution, this has a fractional flux contribution at K-band $f_{\rm AD}$ of $\sim$$0.16 \pm 0.05$ for NGC4151 (\citealt{Kishimoto13KI} and references therein). Therefore, assuming that this component remains unresolved, we calculated the $V^2$ for the dust emission component (quoted in Table~\ref{tab_data}) and the corresponding thin-ring radii. These are the radii shown in Figure~\ref{fig_radius}. 

\begin{figure*}[ht!]
    \centering
\includegraphics[width=\textwidth]{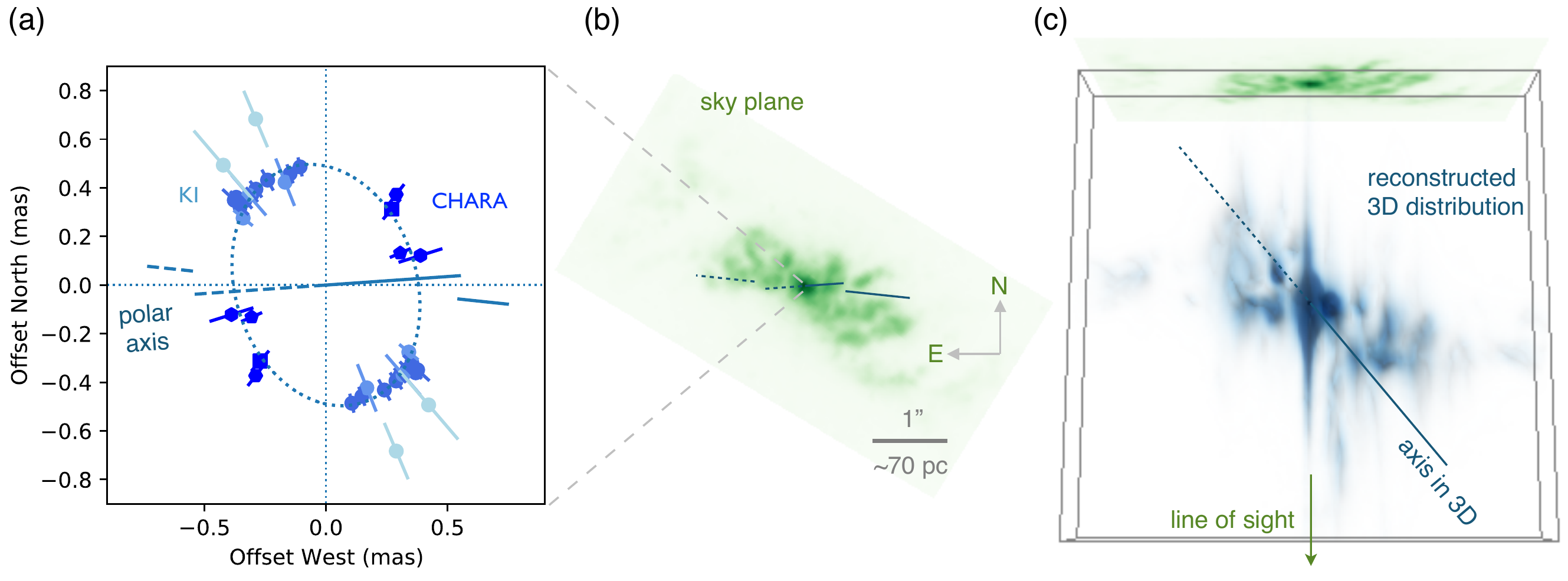}
\caption{($a$) The same ring radii as in Figure~\ref{fig_radius} but with the fitted thin-ring model shown in dotted line. The inner dashed-solid line indicates the polar axis as probed by the optical polarization, and the outer dashed-solid line segments show the PA of the radio jet, exactly the same as in Figure~\ref{fig_radius}.
($b$) HST/WFPC2 archival image of the narrow-line region in [OIII]5007 line, with the two sets of dashed-solid line segments corresponding to those in panel $a$.
($c$) Three-dimensional distribution of [OIII]-emitting clouds reconstructed from HST/STIS multi-slit data, which cover the volume indicated by gray lines. The HST image from panel (b) in green is also shown at the back of the 3D cube for reference, as the 2D flux distribution projected on to the sky plane. This is as seen almost perpendicular to our line of sight. An approximate axis of this 100-pc-scale conical structure is indicated in a dashed-solid line.}
\label{fig_ring_3D}
\end{figure*}

In Table~\ref{tab_fits}, we show the results of the thin-ring fit to the whole $V^2$ data set, both for the simple ring and AD-corrected ring fit (note the difference is only $\sim$10\%). The latter is shown in Figure~\ref{fig_ring_3D}$a$ in dotted line, on top of the same ring radii as shown in Figure~\ref{fig_radius}. The projected ring has its major axis at a PA of 17$\pm$10\degr\, with a radius 0.507$\pm$0.047 mas. The minor to major axis ratio is 0.74$\pm$0.09, thus showing an elongation.

The lack of baseline length coverage between 90 and 190 m leaves some uncertainty in the geometric fit to reproduce the data. An alternative would be a power-law brightness distribution with a certain inner cut-off, a distribution that is more extended in the outer region and more compact in the inner region. However, we found that such a shallow power-law fit to the observed visibilities yields a too small inner cut-off radius, which would be inconsistent with the measured time-lag radius (see section~\ref{sec_intro}). Therefore despite the baseline gap, the adopted thin-ring model seems quite adequate in measuring the overall size and elongation direction of the brightness distribution.

\begin{table}[hb!]
\centering  
 \caption{Thin-ring fit results}
 \begin{tabular}{l c c c c}
  \hline \hline
                  & radius (mas) & min/maj & PA (\degr) \\
  \hline
  all data        & $0.462 \pm 0.050$ & $0.74 \pm 0.10$ & $19 \pm 10$ & \\
  CHARA + 2012 KI & $0.669 \pm 0.086$ & $0.46 \pm 0.08$ & $25 \pm 6$ & \\
  all data with   & \multirow{2}{*}{$0.507 \pm 0.047$}
                  & \multirow{2}{*}{$0.74  \pm 0.09$}
                  & \multirow{2}{*}{$17    \pm 10   $}\\
  $f_{\rm AD}$=0.16$\pm$0.05 & \\
  \hline
 \end{tabular}
 \label{tab_fits}
 For the KI data, we incorporate the conservative systematic uncertainty of 0.03 in $V^2$ (Section 3.2 in \citealt{Kishimoto11}).
\end{table}

\section{Discussions}\label{sec_disc}

\subsection{System axis direction and equatorial elongation}\label{sec_disc_sysaxis}

The system axis direction of the nucleus of NGC4151 can be inferred from a few different observations. At a few arcsec scale, HST images show that the narrow-line region is extended roughly along an E-W or NE-SW direction as shown in Figure~\ref{fig_ring_3D}$b$, indicating that the polar axis of the system is along this direction. A more accurate reference at a slightly smaller spatial scale is the linear jet-like structure observed at radio wavelengths. The one-arcsec-scale radio map clearly shows a linear structure roughly extended along E-W direction, with some systematic wiggles along the structure \citep{Mundell03}. In the innermost region where the component D is considered to have the active nucleus (\citealt{Mundell03,Ulvestad05}), the structure is linearly extended along $\sim$84\degr over $\sim$50 mas scale at $\sim$10 mas resolution \citep{Mundell03}. The brightest sub-component D3 has been shown to be also extended roughly along the same direction at a $\sim$1 mas resolution (\citealt{Ulvestad05}; see section~\ref{sec_relations}).

The reference for the axis direction at an even smaller scale is from optical spectropolarimetry. NGC4151 shows an optical polarization of $\sim$1\%, with its PA parallel to the linear jet direction, which is quite typical for Type 1 AGNs (\citealt{Antonucci83may}). The polarization is believed to originate from electron scattering in an optically-thin, equatorial region (\citealt{Antonucci83may,Smith04may,Lira20}), located somewhere between the broad-line region (or co-spatial with the broad-line region) and dust sublimation region. Thus, in terms of the spatial scale, this is probably the most suitable reference. The polarization seems to be variable, namely 91\degr\ and  97\degr, observed $\sim$2 years apart (1992 / 1994-1995; \citealt{Martel98})

These directions are graphically shown in Figure~\ref{fig_radius} -- the inner dashed-solid line indicates the polarization PA (at 94\degr, the average of the two above), while the outer dashed-solid line segments corresponds to the linear jet direction. The same directions are shown also in Figure~\ref{fig_ring_3D}$a$ and $b$. If we consider a roughly axisymmetric structure, these directions are supposed to indicate the polar axis direction of the system. Therefore, the elongation observed with the near-IR interferometry (thin-ring fit is shown in Figure~\ref{fig_ring_3D}$a$) is approximately perpendicular to this system axis, implying that the near-IR continuum emitting region looks elongated rather along the equatorial direction of the system.

\subsection{Size variability}\label{sec_var}

Due to the variability of the nuclear luminosity, the radius of the dust sublimation region could vary over the years. The radius of the near-IR emitting region in NGC4151 has been claimed to show variation both in the near-IR reverberation \citep{Minezaki04,Koshida09,Schnulle15} and interferometry \citep{Kishimoto13KI}. \cite{Pott10} pointed out no significant increase in interferometric radii against the flux increase by a factor of 2, while \cite{Kishimoto13KI} argued that the radius change might be occurring as a function of the averaged AD flux over several years. On the other hand, \cite{Schnulle15} showed that the time-lag radius decreased as opposed to the interferometric radius increase. If the radius does vary, however, it might not be adequate to compare KI and CHARA data here, since the KI and CHARA observations are apart in time over $\sim$10 years. 

To quantify its potential effect, we have estimated the current expected size based on the prescription for the variability given in \cite{Kishimoto13KI}, where the size is not determined by the instantaneous nuclear flux, but it varies as a function of its long-time average (because the size variation would depend on the timescale of dust sublimation and dust formation). We supplemented the light curve shown in Figure~3 of \cite{Kishimoto13KI} with the monitoring data available from AAVSO\footnote{http://www.aavso.org} for NGC4151 over the last $\sim$10 years, and calculated the suggested 6-year flux average. The flux has been relatively steady since the last high state occurred around 2011, and the resulting predicted size at the current period is calculated to be roughly the same (within $\lesssim 5$\%) as that in 2012. We have then implemented a thin-ring fit to the CHARA data only with the 2012 KI data (i.e. exclude the ones from 2009-2011). The results are shown in Table~\ref{tab_fits}. The minor-to-major axis ratio becomes smaller, but the elongation PA does not change significantly so that the elongation direction still seems roughly perpendicular to the system axis.

\subsection{Inclined ring and inclination angle}\label{sec_inc}

The near-IR continuum emission at $\sim$2 $\mu$m is believed to be predominantly from the innermost, hottest dust grains almost at the sublimation temperature. We tend to consider these sublimating dust grains to be concentrated in the equatorial plane of the system, probably slightly outside the broad-emission-line region. The elongation observed above is consistent with such a naive expectation. 
In this case, we can formally derive the inclination angle from the ellipticity of the elongation, assuming that the region has a ring-like structure. From the minor to major axis ratio of the thin-ring fit described above (Figure~\ref{fig_ring_3D}$a$), we would infer an inclination angle of $\sim$43 $\pm$ 6\degr.

We can compare this value with other estimations from different data sets. Among others, \cite{Das05} deduced an inclination angle of 45$\pm$5\degr\ by modeling the HST/STIS multi-slit data with a conical hollow outflow model over a 100 pc scale region. The modeling is based on the assumption of a “linear” outflow where the velocity of clouds are radial in direction and linearly proportional to the distance from the nucleus. In fact, if we adopt the same velocity field, the distribution of line-emitting material in three-dimensional space can directly be reconstructed from the data, as shown by \cite{Miyauchi20} for a similar HST/STIS data set for the Seyfert 2 galaxy NGC1068. Here we have implemented the same 3D reconstruction using the data set of \cite{Das05} for NGC4151. The result is shown in Figure~\ref{fig_ring_3D}$c$. Details are described elsewhere, but the 3D distribution visualizes and illustrates the possible configuration of the material at a 100 pc scale, or 1 arcsec scale. Consistent with the modeling of \cite{Das05}, it shows a patchy, but approximately hollow and conical distribution, and does suggest an inclination consistent with the value inferred from the assumption of an equatorial ring for the interferometry data. The SW side is the conical outflow toward us, as deduced by \cite{Das05}. It is quite striking that the 100 pc scale structure and the much smaller scale structure probed by our CHARA Array observations suggest similar inclinations.

\begin{figure*}[ht!]
    \centering
\includegraphics[width=\textwidth]{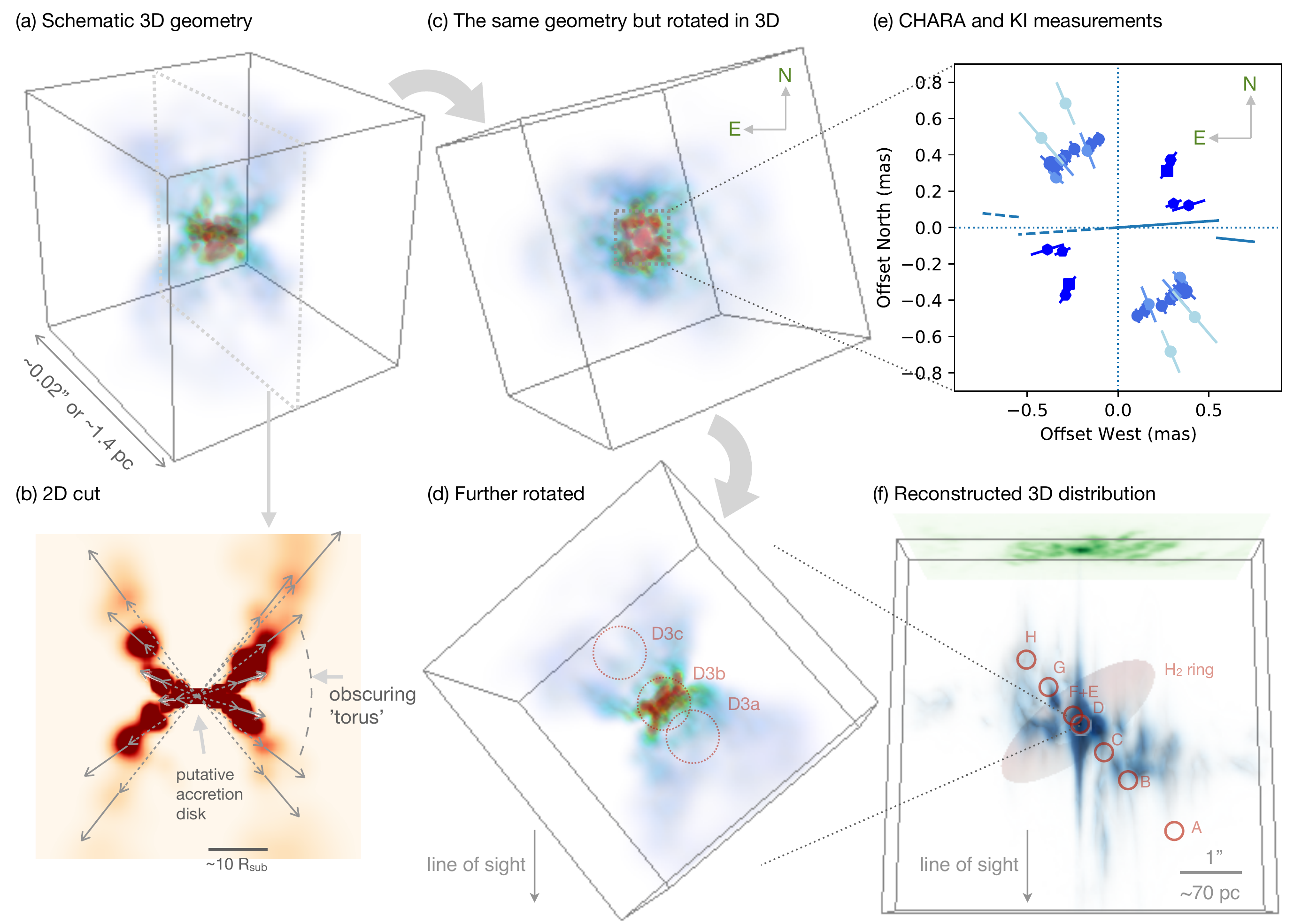}
\caption{
($a$) Schematic 3D geometry inferred for the central region (reproduced from Figure~8b in \citealt{Miyauchi20}). 
($b$) 2D cut of the 3D geometry shown in panel $a$ (reproduced from Figure~8d in \citealt{Miyauchi20} with additional annotations).
($c$) The same geometry as in panel $a$, but rotated in 3D to match the possible configuration in NGC4151 as seen from us, where the central dotted square schematically corresponds to the region mapped by the CHARA Array and KI as indicated in panel $e$, which is the same as Figure~\ref{fig_radius}.
($d$) The same 3D geometry but further rotated to a side view, perpendicular to our line of sight to the object. Approximate 3D locations inferred for the VLBI components (15 GHz; \citealt{Ulvestad05}) are schematically shown in red circles.
($e$) See the description for panel $c$.
($f$) Reconstructed 3D distribution shown in Figure~\ref{fig_ring_3D}c, but 
with approximate 3D positions inferred for the knots A--H of the radio jet (\citealt{Mundell03}). The molecular hydrogen distribution as inferred by \cite{Fernandez99} is shown as a red transparent ring in 3D with inner and outer radii of 0.5 and 1.5 arcsec.
}
\label{fig_schematic}
\end{figure*}

\subsection{Why in an equatorial plane?}\label{sec_whyeq}

In the mid-IR, long-baseline interferometry has shown that many objects have polar-elongated morphology at pc scales, or tens of $\Rsub$ scales where $\Rsub$ is the dust sublimation radius (\citealt{Hoenig12,Hoenig13,Tristram14,LopezGonzaga14,LopezGonzaga16}), though the mid-IR interferometry data for NGC4151 are not clear about this point, since the PA coverage of the data is poor (\citealt{Kishimoto11MIDI}; \citealt{Burtscher13}). On the other hand, the mid-IR interferometry for large angular size AGNs (i.e. NGC1068 and Circinus galaxy) indicates an equatorial elongation in several $\Rsub$ scales, embedded in the polar elongation at much larger scales of tens of $\Rsub$ (\citealt{Raban09,Tristram14,LopezGonzaga14}). This has now even more clearly been shown by mid-IR interferometric imaging for NGC1068 (\citealt{GamezRosas22}; see more below) and for Circinus (\citealt{Isbell22}). Since we are probing a few $\Rsub$ scale for NGC4151 here ($\lambda/b \sim 1.8$ mas; $\Rsub$ $\sim$0.3$-$0.5 mas for NGC4151), our detection of the equatorial elongation seems geometrically consistent with such a structure seen by the mid-IR interferometry.

In fact, with the near-IR interferometer VLTI/GRAVITY, \cite{Leftley21} has detected an equatorial elongation in the near-IR for the Type~1 AGN ESO323-G77, despite the small visibility decrease at baselines up to 130~m. The object shows a polar elongation in the mid-IR \citep{Leftley18}. Thus the equatorial elongation as seen by the CHARA Array for the type 1 AGN NGC4151 seems to fit in with all these interferometric results.

The polar-elongated morphology found in the mid-IR at the pc scale, combined with the larger-scale HST-resolved outflow, would suggest a conical and hollow outflowing structure also at the inner pc scale. This wind has been suggested to be driven by the radiation pressure on dust grains \citep{Hoenig12,Hoenig13}. In this case, it is likely to be launched at the dust sublimation region. There, the dusty gas could be inflated or puffed up by the infrared radiation of the heated grains \citep{Krolik07}, then radially accelerated by the UV radiation from the nucleus and form the hollow conical structure \citep{Hoenig19}. This near-IR emitting, wind-launching region might seem a distinct, separate component in the spectral energy distribution (e.g. \citealt{Mor09, Kishimoto11MIDI,Hoenig12}), and in particular, this could correspond to the bump-like feature at around 3 $\mu$m seen in the spectra of many Type 1 AGNs (\citealt{Hoenig17}). Regarding this near-IR emitting region as a slightly extended disk and the mid-IR emitting component as a dusty wind, the model of \cite{Hoenig17} has successfully explained the near-IR and mid-IR observations of the type 1 AGNs NGC3783 and ESO323-G77 including interferometry (\citealt{Leftley21}).

However, the observed emission properties of the 100-pc-scale outflow, the pc-scale polar elongation, and the sub-pc-scale equatorially-elongated structure, might actually suggest that all these three originate from the same nuclear wind. They are all rather continuously distributed on the plane of the surface-brightness and color temperature (Figure~7 in \citealt{Miyauchi20}). We can also estimate the infrared emissivity (i.e. surface brightness in units of Planck function) of these components on this plane, and infer the optical thickness. While the 100 pc scale outflow is UV-optically thin, the pc-scale structure (showing the polar elongation in the mid-IR) seems UV-optically thick. This means that this inner hollow outflow would provide the obscuration of the nucleus in equatorial directions in the UV/optical, as also argued by \cite{Hoenig19}. In this case, the whole outflow would be the obscuring ‘torus’, and we would not need to invoke the disk component for obscuration described above. The near-IR emitting component would not be a distinct one, but would simply be the innermost part of the outflow. But then, we would rather expect that it would look more polar-elongated, as opposed to the CHARA Array results presented here.

\subsection{Flaring geometry from episodic, one-off acceleration}

One consistent picture might be that the distribution actually is geometrically flaring effectively (see Figure~\ref{fig_schematic}a), with the cloud acceleration being one-off, or episodic, rather than continuous. The geometrical flaring would give an inner-equatorial, outer-polar structure, explaining the mid-IR interferometry results for NGC1068 and Circinus described above with a single structure (i.e. explains the two components of the inner-equatorial and outer-polar structures as a single entity) and the CHARA Array results at the same time. Here, we do not infer that each cloud follows the flaring path, as it would need an additional acceleration toward the polar-axis direction in the outer radii. Rather, in fact, it has been argued that the velocities of the outflowing clouds at 100 pc scales could be from a one-off, episodic acceleration (once in $\gtrsim$10$^5$ yr) at the innermost wind-launching site, since they are Hubble-flow-like, i.e. the velocity of each cloud seems proportional to its distance to the nucleus (\citealt{Ozaki09, Miyauchi20}). The geometrical flaring above can realize if each cloud is simply moving radially roughly at a constant speed but the clouds have a polar-fast, equatorially-slow velocity distribution from the one-off anisotropic acceleration (see Figure~\ref{fig_schematic}b), which might be reasonable for the radiation pressure from an accretion disk.

This possible episodic acceleration might actually be quite consistent with 
the episodic activity that has been inferred from X-ray observations. \cite{Wang10} has identified a significant soft X-ray emission extended out to $\sim$$10^4$ light years from the nucleus of NGC4151. This emission has been interpreted to indicate either mechanical heating by an episodic outflow or photoionization by a highly-luminous, Eddington-limited outburst. In both cases, the episodic activity is estimated to have occurred in the last $10^4$-$10^5$ years, which is a timescale  quite similar to what we discuss here.

The flaring geometry from an episodic acceleration (Figure~\ref{fig_schematic}a) as seen edge-on might fit with the mid-IR interferometric image in \cite{GamezRosas22} (their Figure~1e, mid-IR RGB-composite image). The illustration in Figure~\ref{fig_schematic}a, which is from Figure~8 in \cite{Miyauchi20},  intends to represent the distribution of the ionized gas and $T \gtrsim300$ K dust, and the inner equatorial region (represented in red) would correspond to the near-IR emitting, dust sublimation region. This might also be consistent with the 3.7 $\mu$m image in \cite{GamezRosas22} showing an equatorial structure, with the dust sublimation region being still hidden below. In the case of the Type 1 object NGC4151, we would see the sublimation region directly --- Figure~\ref{fig_schematic}c shows the same 3D illustration, but rotated in 3D to match the inferred inclination and PA of the axis projected on the sky for NGC4151  with the central square schematically corresponding to the CHARA Array and KI observations (Figure~\ref{fig_schematic}e). This illustrates the possible configuration in NGC4151. Figure~\ref{fig_schematic}d shows the same geometry further rotated to have a view perpendicular to both our line of sight and the system axis, corresponding to the view of the reconstructed 3D distribution at a much larger scale shown in Figure~\ref{fig_ring_3D}c and reproduced in Figure~\ref{fig_schematic}f.

\subsection{Relations to other observations and simulations}\label{sec_relations}

How would this structure be related to those observed at other wavelengths and other scales? As mentioned in section \ref{sec_disc_sysaxis}, the radio continuum source D3, believed to contain the nucleus (\citealt{Mundell03}), has been resolved at $\sim$1 mas resolution with VLBI (a beam size of 0.95 $\times$ 0.46 mas; \citealt{Ulvestad05}). The source is resolved into three components along the E-W radio jet direction. The middle component D3b is argued to be the nucleus, with the other two knots $\sim$1$-$2 mas away from it. Thus the dust sublimation region of $\sim$0.5 mas in radius, as resolved by the CHARA Array observations here, would be located at the center of the three-knot jet-like structure that has a scale comparable to that of the sublimation region itself. This is schematically shown in Figure~\ref{fig_schematic}d, where approximate 3D positions of the three knots are indicated. Here the jet axis at this scale is assumed to coincide with the system axis, based on the inclination of $\sim$40\degr\ inferred for the radio jet at larger scales (\citealt{Pedlar93}) which is consistent with the inclination of the system discussed in section~\ref{sec_inc}.

At slightly larger, 10--100 mas scales, neutral hydrogen absorption is observed in one of the several components of the linear radio jet. This is the component called E+F, which is at the counter-jet side but closest to the radio nucleus D3 (the projected distance is $\sim$70 mas or $\sim$5 pc; \citealt{Mundell95, Mundell03}). In Figure~\ref{fig_schematic}f, we have drawn approximate 3D positions of the jet components assuming again the inclination of 40\degr. In terms of the episodic polar-fast outflow discussed above, this foreground neutral hydrogen material in front of the counter jet component E+F might be a part of the slow-velocity outflows moving relatively close to equatorial directions. In the scenario of the one-off acceleration $\sim$10$^5$ years ago, the clouds several pc away from the nucleus would be at the velocity of several 10s of km/s, which would be consistent with the observed radial velocities of the absorption lines (Figure~5b of \citealt{Mundell03}). 

This central region seems to be surrounded by a molecular ring-like structure at slightly outer radii of $\sim$1 arcsec, which is observed in the infrared molecular hydrogen line H$_2$ 1-0 S(1), with the projected major axis PA $\sim$170\degr\ and inclined by $\sim$37$\degr$ from our line of sight (\citealt{Fernandez99}). This configuration is schematically shown in 3D in Figure~\ref{fig_schematic}f as a red transparent disk (inner and outer radius of 0.5 and 1.5 arcsec, respectively). The kinematics of this ring is inferred to be dominated by rotation, rather than outflow, based on their Fabry-Perot image missing the North-South part which would possess higher radial velocities if in rotation. Therefore, this ring might be kinematically distinct from the obscuring and outflowing torus discussed above, and rather be an outer part of the ‘fueling disk’, an equatorial thin disk-like structure that transfers material toward the nucleus (\citealt{Hoenig19}; \citealt{Miyauchi20}).

The idea to have a wind responsible for the obscuration of the nucleus has been investigated with theoretical calculations and numerical simulations. The obscuring wind could be from a combination of X-ray heating and radiation pressure (\citealt{Krolik86, Krolik88, Wada12}), or, as mentioned in section~\ref{sec_whyeq}, triggered by IR radiation pressure from heated dust and accelerated by the AGN UV radiation (\citealt{Krolik07, Chan16, Chan17, Dorodnitsyn16, Namekata16, Williamson19, Williamson20}).
As for the episodic property, the hydrodynamical winds simulated by \cite{Wada12,Wada15} are in fact quite intermittent in nature.
Here we argued observationally that an episodic wind accelerated by polar-strong radiation, forming an effectively flaring geometry, would be able to explain the near-IR equatorial elongation at a $\Rsub$ scale, mid-IR polar elongation at tens of $\Rsub$ scale, as well as the linear velocity field and the hollow conical outflow at a 100 pc scale, at least qualitatively. 
While the exact physical processes for launching winds are still being explored theoretically, a modeling focused on reproducing these observed properties including possible episodic acceleration would enable us to do more quantitative comparisons with observations.
We plan to implement such modeling as well as acquire further data on this innermost structure with the CHARA Array.

\section{Summary and conclusions}

We have presented the near-IR interferometric measurements of the Type 1 AGN NGC4151 at projected baselines up to $\sim$250 m, the longest baseline achieved so far in the infrared for an extragalactic object. With these baselines, the compact near-IR structure, or the dust sublimation region, has now been spatially resolved. The measurements provide the size along different position angles, giving the first information on the morphology of the object at this inner spatial scale. When combined with the previous Keck interferometer data albeit at shorter baselines, we find that the region is elongated along roughly north-south, perpendicular to the system axis direction as determined by the linear radio jet structure and optical polarization observations. 

We argued that an effectively flaring geometry can explain both the inner-equatorial and outer-polar elongation simultaneously, which are seen in our CHARA Array observations and the mid-IR interferometry. The flaring geometry at a given time could be formed from a polar-fast, equatorially-slow, one-off acceleration, while the episodic acceleration is quite consistent with the Hubble-flow-like velocity field observed at a $\sim$100 pc scale. Long-baseline measurements covering a wide position-angle range in a single epoch are now desired to confirm the results. We are now aiming for acquiring such data at the CHARA Array, which is currently the only infrared interferometer that has such long baselines and sufficient uv coverage.

\acknowledgements

This work is based upon observations obtained with the Georgia State University Center for High Angular Resolution Astronomy Array at Mount Wilson Observatory.  The CHARA Array is supported by the National Science Foundation under Grant No. AST-1636624 and AST-2034336.  Institutional support has been provided from the GSU College of Arts and Sciences and the GSU Office of the Vice President for Research and Economic Development. We are grateful for all the generous and patient support from the CHARA staffs since the start of this project in 2010 at the CHARA Array. A part of the time at the CHARA Array was granted through the NOIRLab community access program (NOIRLab PropIDs: 2010A-0081, 2012A-0187, 2013A-0241, 2014A-0072, 2015A-0242, 2016A-0091, 2017A-0049, 2018B-0109, 2019A-0055, 2021A-0013; PI: M. Kishimoto). This work was supported in part by JSPS grants 16H05731, 20K04029 and 21H04496, and the grant E1906 through Kyoto Sangyo University. We acknowledge with thanks the variable star observations from the AAVSO International Database contributed by observers world-wide and used in this research. This research has made use of the Jean-Marie Mariotti Center Aspro and SearchCal services.


\appendix

\section{Observed raw visibilities and power spectra of fringe tracks}

\begin{figure*}[t!]
\centering
\includegraphics{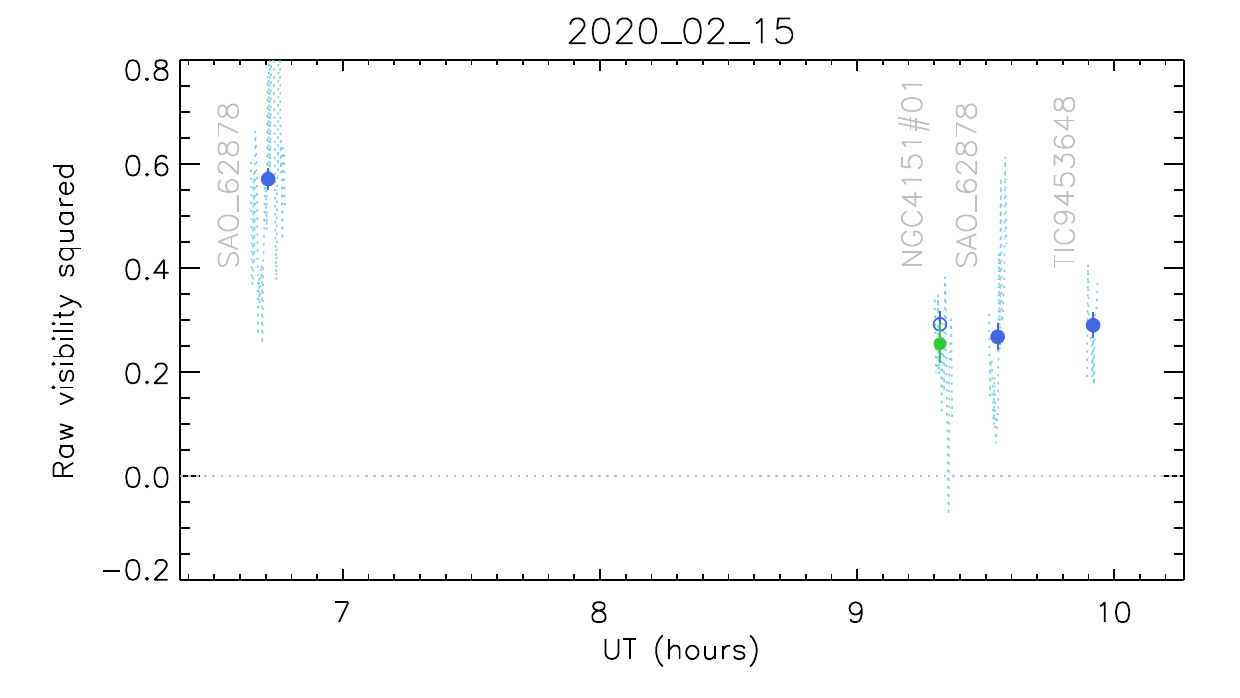}
\caption{Observed raw visibilities in the night of 2020-02-15 at CHARA. 
Light dotted lines show the raw visibilities averaged over 10 scans ($\sim$15 seconds) in each track of objects. Blue filled circles show the raw visibilities integrated over each track of calibrators with uncertainties. Blue open circles are the system visibilities estimated from the calibrator observations. The green filled circle is the raw visibility of the science target from the integration over the track.
}
\label{fig_night_2020_02_15_night}
\includegraphics[width=\textwidth]{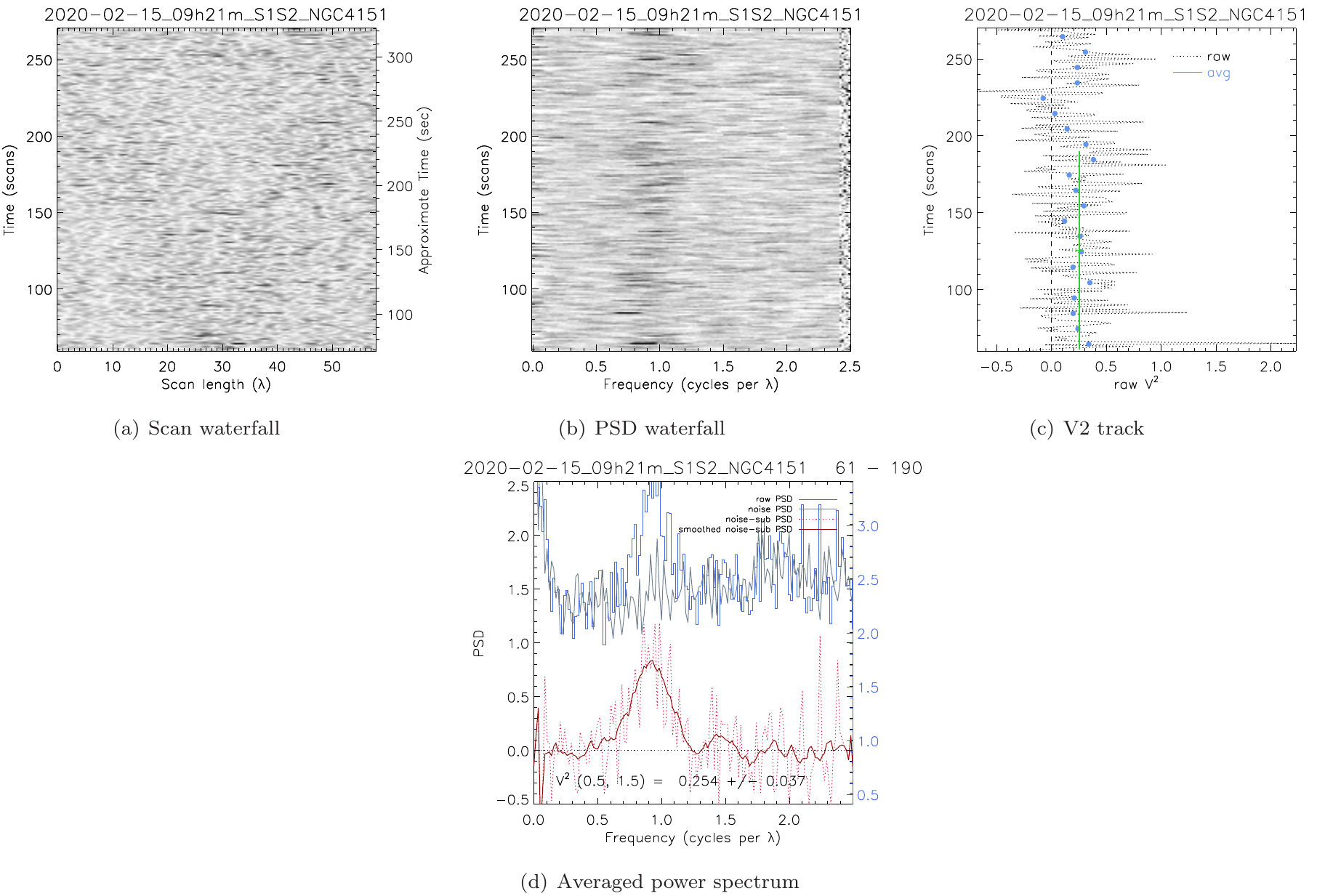}
\caption{Details of the fringe track for NGC4151 on 2020-02-15. (a) Fringe envelopes of the scans are shown in grayscale, with each scan in horizontal direction, forming a waterfall plot along the vertical direction. (b) Power spectrum of each scan is shown in grayscale along the horizontal direction, again forming a waterfall plot of power spectra. (c) Dotted lines show the raw $V^2$ obtained from integrating each power spectrum, and the blue dots indicate its 10-scan average. (d) Averaged power spectrum is shown in red dotted line, with its smoothed version in red solid line. The blue spectrum is before the subtraction of the noise bias, which is shown in gray, with the scale axis at the right side in the same blue.}
\label{fig_night_2020_02_15_track01}
\end{figure*}

Figure~\ref{fig_night_2020_02_15_night} shows the raw visibilities for each calibrator and the target observed in the night of 2020-02-015. For each object, the raw visibilities obtained by averaging the power spectra over 10 scans (corresponding to an integration over $\sim$15 seconds) are shown in dotted lines. The visibilities integrated over each track of calibrator observations are shown in filled blue circles with uncertainties. The open circles show the system visibilities (transfer functions) at the time of the calibrator observations (almost overlapping with the calibrators’ raw visibilities) and at the time of the science target observation. The latter is obtained from the linear interpolation between the system visibilities bracketing the science target observation. The visibility integrated over the track (or a portion of the track; see below) for the science target is shown in a green filled circle with its uncertainty. 

Figure~\ref{fig_night_2020_02_15_track01} shows the details of the fringe track for NGC4151 on 2020-02-15. The top-left panel (a) shows the envelope (fringe amplitude) over each scan in grayscale (note the right-side axis showing approximate elapsed time in seconds). The top-middle panel (b) shows the corresponding power spectra also in grayscale. Fringes should be detected at the frequency of 1 cycle per observing wavelength $\lambda$. They are clearly seen with some wiggles around this frequency over different scans. The top-right panel (c) shows the raw $V^2$ value for each scan in dotted lines, while the light blue dots correspond to the averages over 10 scans. In the bottom panel (d), the red dotted line is the final averaged power spectrum, with the same spectrum smoothed over 10 frequency pixels shown in red solid line. This is after the subtraction of the noise bias spectrum, which is measured from the average of the same number of off-fringe scans as the object scans. Here the power spectrum before the subtraction is shown in blue and the noise bias spectrum in gray, with the scale axis for both indicated at the right side in blue.

In this particular fringe track, the fringes slipped away from the scan range at and around the scan \#220. Therefore, we integrated the power spectra until scan \#190. The integration range is indicated with a green vertical line in panel (c), with its x-axis position corresponding to the value of the obtained $V^2$ from the integration. The uncertainty is also shown in the horizontal green line at the mid-point of integration. Note that the exact choice of the scan number for integration does not alter the results beyond the estimated uncertainty.

\begin{figure*}[t!]
\centering
\includegraphics{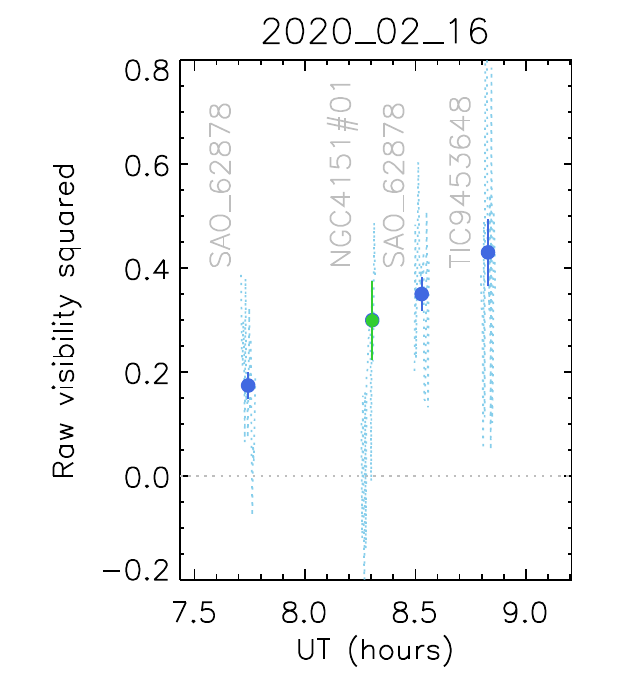}
\caption{Observed raw visibilities in the night of 2020-02-16 at CHARA. The notations are the same as in Figure~\ref{fig_night_2020_02_15_night}.}
\label{fig_night_2020_02_16_night}
\includegraphics[width=\textwidth]{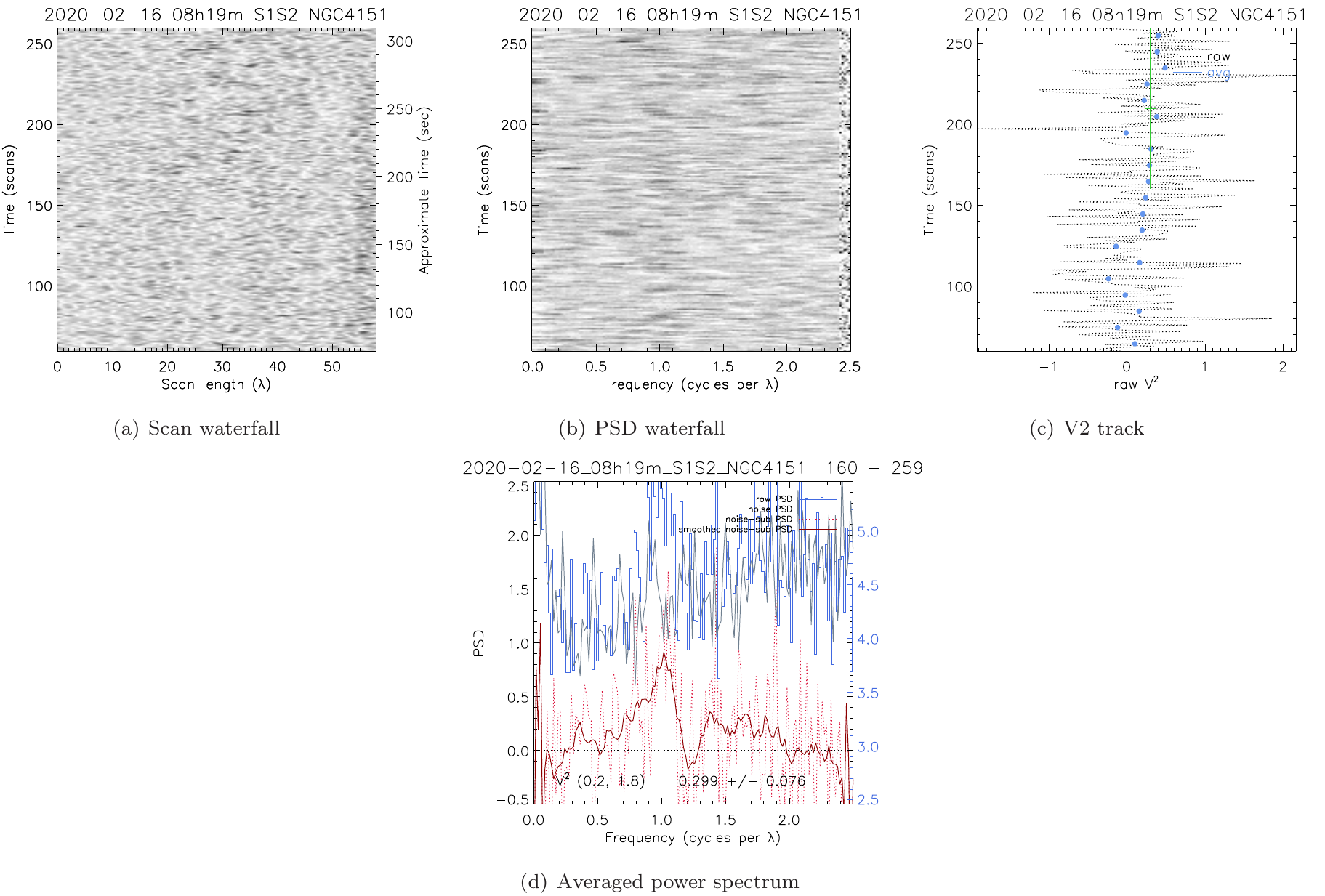}
\caption{Details of the fringe track for NGC4151 on 2020-02-16, in the same format as in Figure~\ref{fig_night_2020_02_15_track01}.}
\label{fig_night_2020_02_16_track01}
\end{figure*}

\begin{figure*}[t!]
\centering
\includegraphics{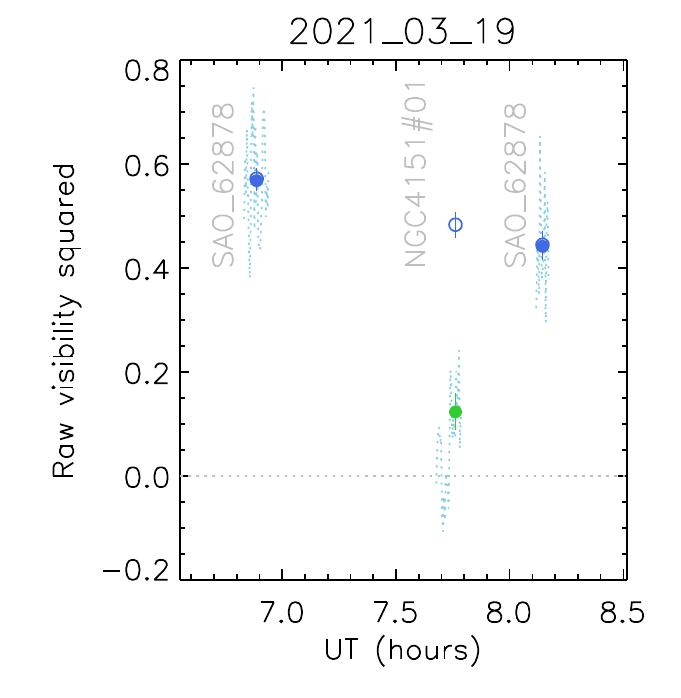}
\caption{Observed raw visibilities in the night of 2021-03-19 at CHARA. The notations are the same as in Figure~\ref{fig_night_2020_02_15_night}.}
\label{fig_night_2021_03_19_night}
\includegraphics[width=\textwidth]{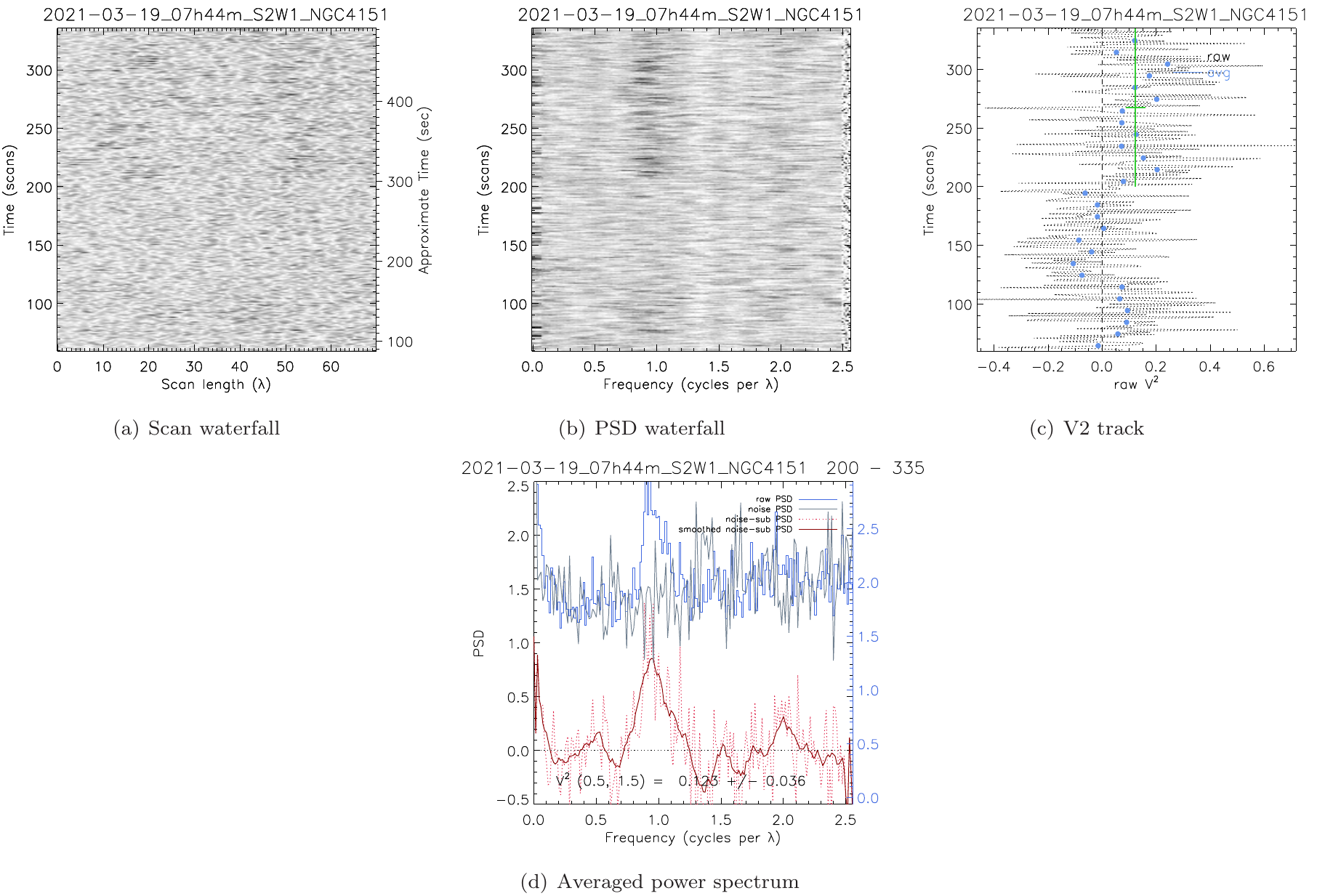}
\caption{Details of the fringe track for NGC4151 on 2021-03-19, in the same format as in Figure~\ref{fig_night_2020_02_15_track01}.}
\label{fig_night_2021_03_19_track01}
\end{figure*}

\begin{figure*}[t!]
\centering
\includegraphics[width=\textwidth]{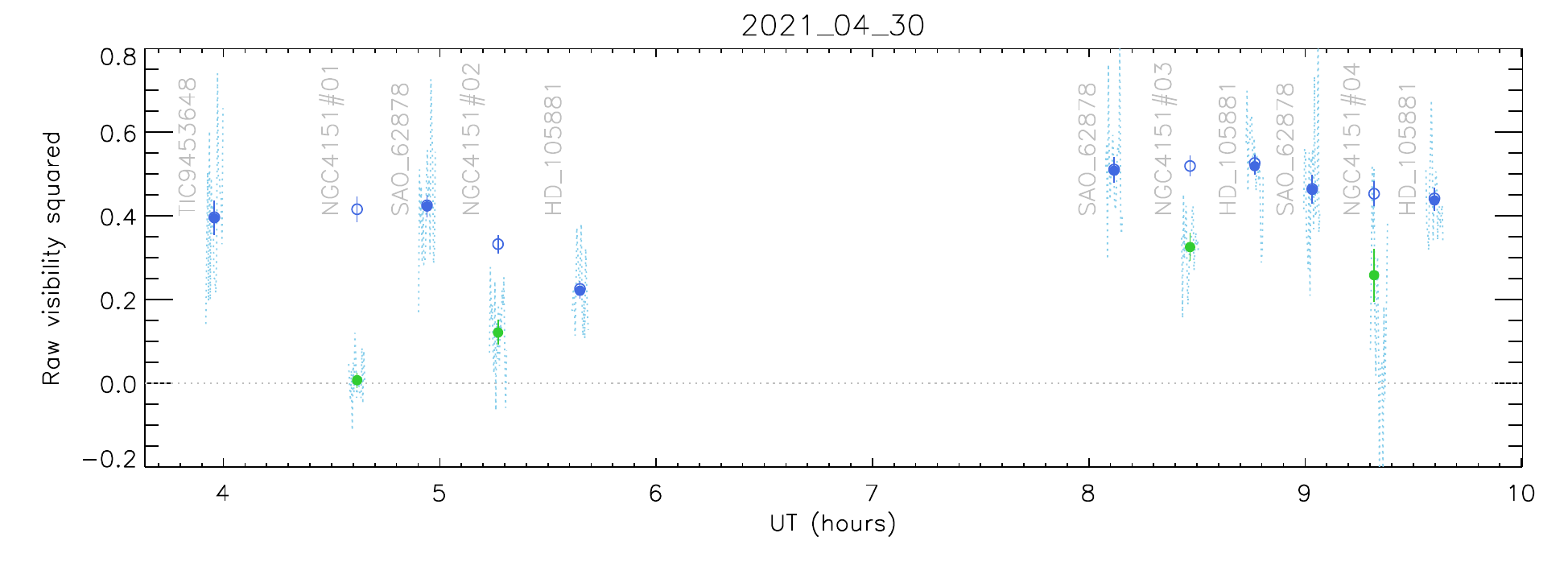}
\caption{Observed raw visibilities in the night of 2021-04-30 at CHARA. The notations are the same as in Figure~\ref{fig_night_2020_02_15_night}.}
\label{fig_night_2021_04_30_night}
\end{figure*}

\begin{figure*}[h!]
\includegraphics[width=\textwidth]{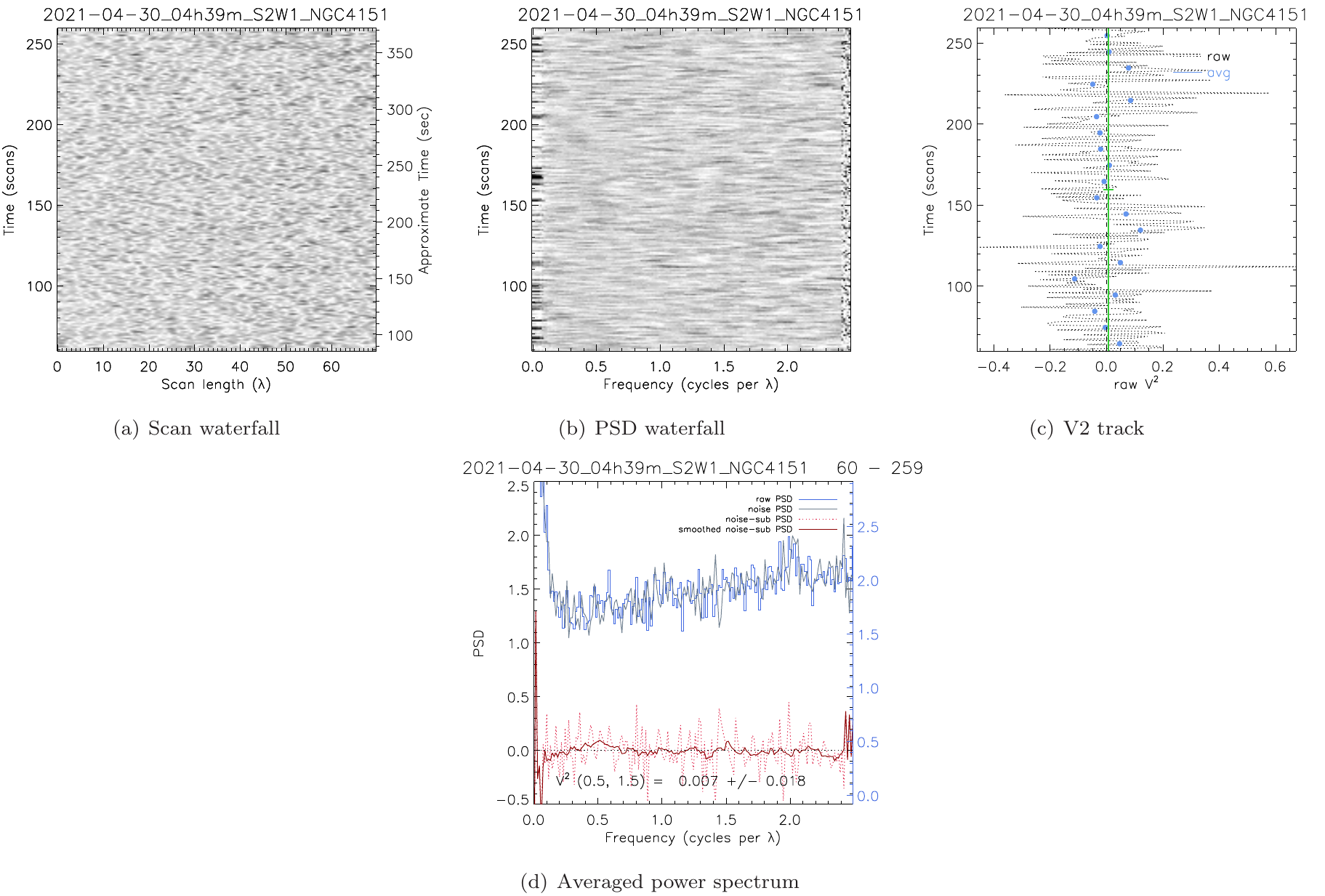}
\caption{Details of the fringe track \#01 for NGC4151 on 2021-04-30, in the same format as in Figure~\ref{fig_night_2020_02_15_track01}.}
\label{fig_night_2021_04_30_track01}
\end{figure*}

\begin{figure*}[h!]
\includegraphics[width=\textwidth]{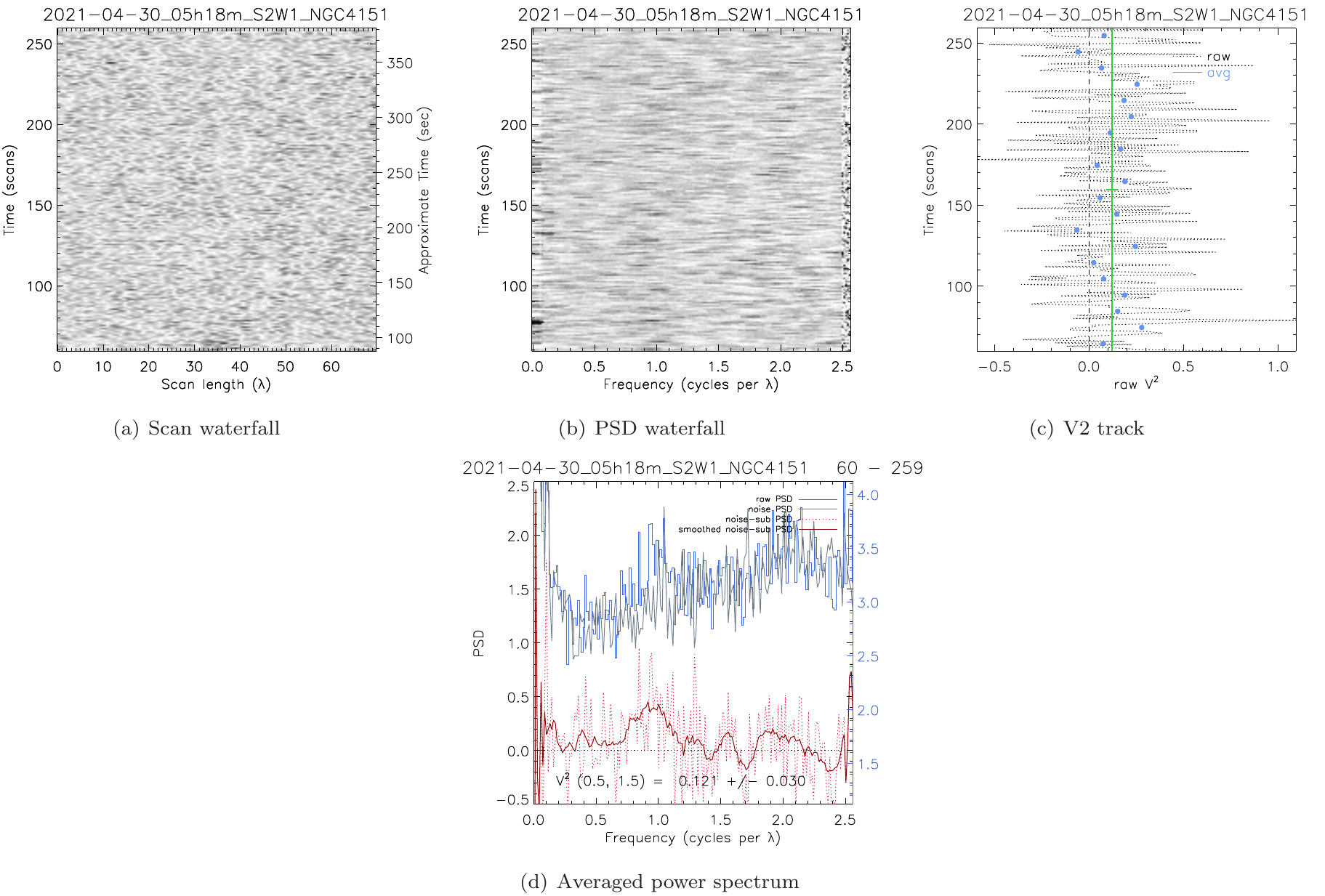}
\caption{Details of the fringe track \#02 for NGC4151 on 2021-04-30, in the same format as in Figure~\ref{fig_night_2020_02_15_track01}.}
\label{fig_night_2021_04_30_track02}
\end{figure*}

\begin{figure*}[h!]
\includegraphics[width=\textwidth]{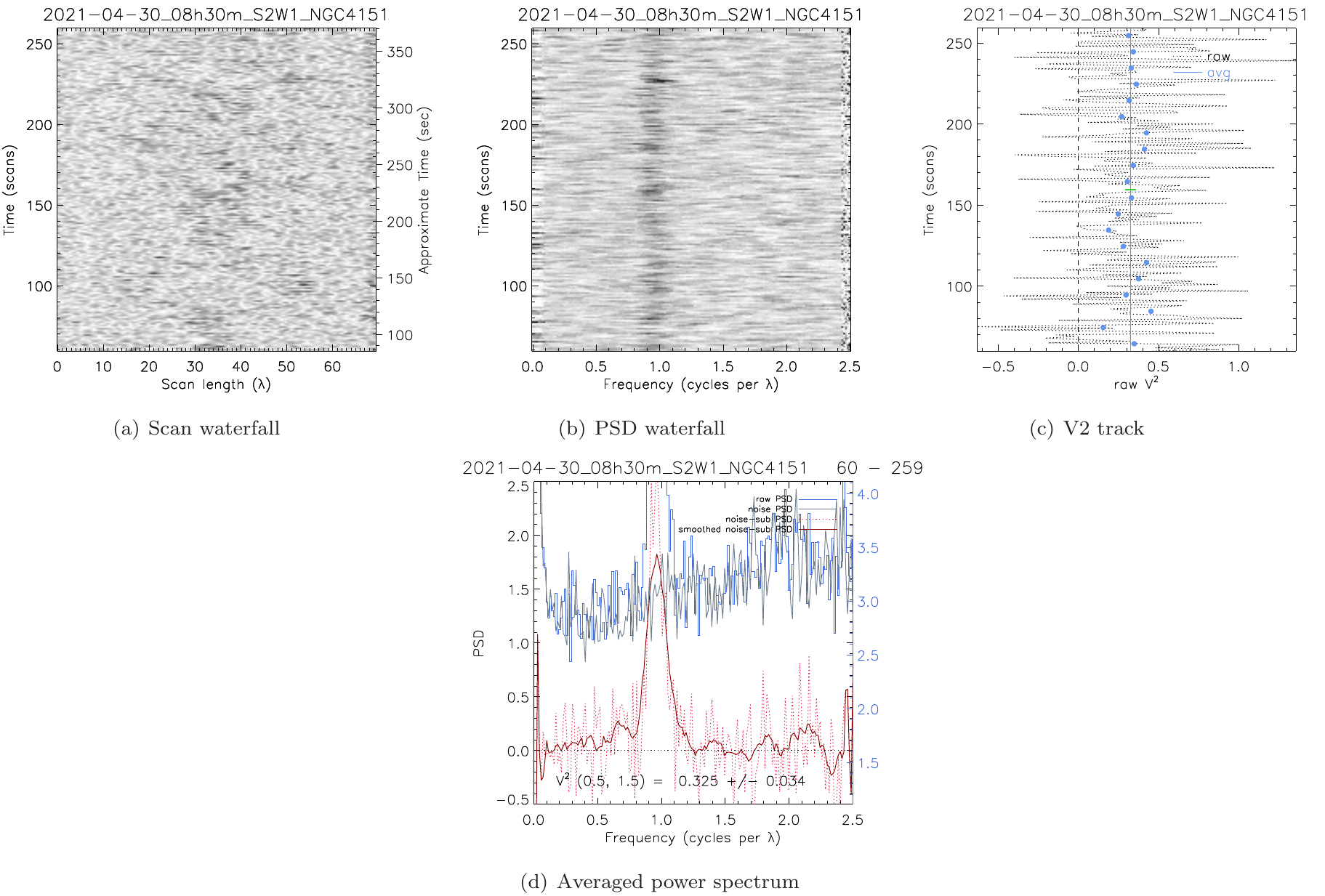}
\caption{Details of the fringe track \#03 for NGC4151 on 2021-04-30, in the same format as in Figure~\ref{fig_night_2020_02_15_track01}.}
\label{fig_night_2021_04_30_track03}
\end{figure*}

\begin{figure*}[h!]
\includegraphics[width=\textwidth]{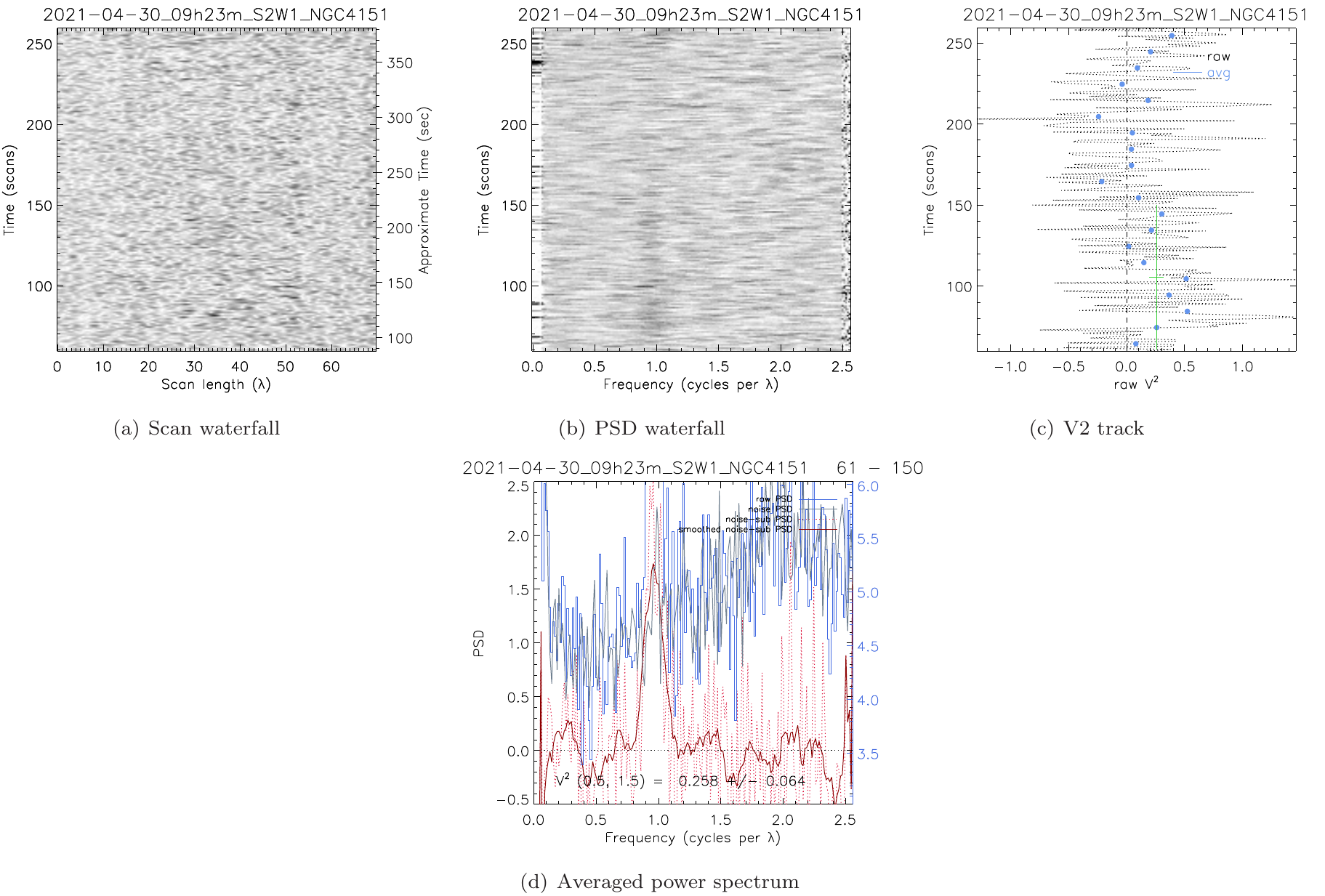}
\caption{Details of the fringe track \#04 for NGC4151 on 2021-04-30, in the same format as in Figure~\ref{fig_night_2020_02_15_track01}.}
\label{fig_night_2021_04_30_track04}
\end{figure*}

Figures~\ref{fig_night_2020_02_16_night}-\ref{fig_night_2021_04_30_track04} show the observed raw visibilities in the same way. Multiple tracks of the science target within a given night are marked with sequential numbers, and shown individually in separate Figures. Note that we did not see fringes in the track \#01 for NGC4151 in the night of 2021-04-30. Since we are not entirely sure if the scans were in the right range, this track was excluded from the analysis, but shown here for completeness.


\end{document}